\documentclass[a4paper]{jpconf}
\usepackage{graphicx}
\usepackage{float} 
\usepackage{xurl} 

\usepackage[margin=1in]{geometry}
\usepackage{hyperref}
\hypersetup{
    colorlinks=true,
    citecolor=blue,
    linkcolor=blue,
    filecolor=blue,      
    urlcolor=blue,
    }
\renewcommand\theenumi{\arabic{enumi}}

\begin{document}

\title{Particle Physics Outreach to K-12 Schools and Opportunities in Undergraduate Education}

\author{Marge G. Bardeen$^1$, Olivia M. Bitter$^2$, Marla Glover$^3$, Sijbrand J. de Jong$^4$, Tiffany R. Lewis$^5$, Michael Fetsko$^6$, Adam LaMee$^7$, Christian Rosenzweig$^8$; Deborah Roudebush$^9$, Andrew D. Santos$^{10}$, Shane Wood$^{11}$, Kenneth Cecire$^{12}$, Randal Ruchti$^{12}$, Guillermo Fidalgo$^{13}$, Sudhir Malik$^{13}$}

\address{$^1$Education Office, Fermilab (Retired), Batavia, IL 60510, USA}
\address{$^2$Physical Science Division, The University of Chicago, Chicago, IL 60637, USA and Neutrino Division, Fermi National Accelerator Laboratory, Batavia, IL 60510, USA}
\address{$^3$Curriculum and Instruction Department, Purdue University, West Lafayette, IN 47907, USA}
\address{$^4$High Energy Physics Department, Institute for Mathematics, Astrophysics and Particle Physics, Faculty of Science, Radboud University, Nijmegen, The Netherlands}
\address{$^5$NASA Postdoctoral Program Fellow, Astroparticle Physics Laboratory, NASA Goddard Space Flight Center, Greenbelt, MD 20771}
\address{$^6$Science Department, Mills E. Godwin High School, Henrico, VA 23238, USA}
\address{$^7$Physics Department, University of Central Florida, Orlando, FL 32816, USA}
\address{$^8$Science Department, Elmwood Park High School, Elmwood Park, IL 60707, USA}
\address{$^9$Fairfax County Public Schools (Retired) 4410 Mariner Lane Fairfax, VA 22033, USA}
\address{$^{10}$Leprince-Ringuet Laboratory, École Polytechnique, 91120 Palaiseau, France}
\address{$^{11}$Mounds View Schools in Shoreview, MN 55126, USA}
\address{$^{12}$Physics Department, University of Notre Dame, IN 46556, USA}
\address{$^{13}$Physics Department, University of Puerto Rico Mayaguez, PR 00682, USA}
\ead{sudhir.malik@upr.edu}

\begin{abstract}
To develop an increase in societal interest in the fundamental sciences of particle physics and particularly for maintaining the support structures needed to succeed in experiments that take several decades to develop and complete, requires strong educational back-grounding at all levels of the instructional system and notably at early stages in the process. While many (particularly young) students might show an early interest and aptitude for science and mathematics at the elementary level, the structures are not necessarily in place to capture, nurture and develop such nascent interests. To encourage and strengthen such interests, strong connections must be made at K-12 and Undergraduate levels. The paper discusses the on-going efforts and makes recommendations.
\end{abstract}

\section{Introduction}

The next generation of science-savvy citizens, physics postdocs, engineers, technicians and science policy-makers are in middle school today. It is imperative to the continued excellence of particle physics that we help build the K-12 STEM pipeline today. Students in high school need to learn about 21st century physics and the role it plays in society in order to be informed voters, science activists, and potential scientists~\cite{k12_CP2B2,k12_CP2B3,k12_CP2B4,k12_CP2B6}. A fundamental component to any successful program will be a goal of equity and inclusion in which young people have access to information and opportunities in physics~\cite{k12_CP2B7,k12_CP2B8,k12_CP2B9,k12_CP2B10}.
To develop an increase in societal interest in the fundamental sciences of particle physics and particularly for maintaining the support structures needed to succeed in experiments that take several decades to develop and complete, requires strong educational back-grounding at all levels of the instructional system and notably at early stages in the process \cite{k12_CP2B1,k12_CP2B5}. While many (particularly young) students might show an early interest and aptitude for science and mathematics at the elementary level, the structures are not necessarily in place to capture, nurture and develop such nascent interests. To encourage and strengthen such interests, strong connections must be made at K-12 and Undergraduate levels. The paper discusses the on-going efforts and makes recommendations.

\section{Outreach to K-12 Schools}

Particle physicists, alone, cannot restructure or change the purpose of the high school physics curriculum that is set by textbook publishers and state adopters such as Texas and California. However, the particle physics community does have a shared responsibility to help teachers and students learn about physics insights into the world around us, how science is done and how to “follow the science” on societal issues. The next generation of science-savvy citizens and post-docs, technicians, and engineers are in middle school today. How can we help build the K-12 STEM pipeline to enable high school students to take physics and learn about topics in 21st century physics and the role physics plays in society? How can we develop a more diverse population of young physics students?

While physicists may engage students directly, the particle physics community is most effective and gets the most from its efforts when working with teachers. Teachers engage more students more deeply than we could ever reach in single events, however interesting or impactful~\cite{k12_CP2C1}. The vast majority of teachers have never done scientific research~\cite{k12_CP2C3}. Like the general public, they learn about scientific discoveries in news accounts that rarely describe the lengthy story from questions and ideas to experiments—designing and building instruments, collecting and analyzing data, publishing results—and only then discovery. And some teachers rely on cookbook labs because they are uncomfortable letting their students do open-ended investigations. At our best, we can help teachers learn to apply the discovery nature of our science in their classrooms~\cite{k12_CP2C7}. Teachers can build the knowledge and skills of their students to use scientific habits of mind, engage in scientific inquiry and reason in a scientific context. Teachers can use the compelling nature of particle physics discovery to motivate and excite students about the study of physics.

We have selected a few programs and resources to highlight in our report. In addition, in 2013, Snowmass on the Mississippi included outreach activities that remain very active with resources online today:
\begin{enumerate}
\item Contemporary Physics Education Project (CPEP) - fundamental particles and interactions, https://www.cpepphysics.org/particles.html, a non-profit organization of teachers, educators and physicists
\item International Particle Physics Outreach (IPPOG) - https://ippog.org an outreach network with a large resource database.
\item The Particle Adventure -  https://particleadventure.org, an award-winning interactive tour from the Particle Data Group of Lawrence Berkeley Laboratory
\item Quarked - http://www.quarked.org, Adventures in the Subatomic World for young children from the University of Kansas
\item QuarkNet - https://quarknet.org, a collaboration of physicists and physics teachers at over 50 centers at universities and labs across the U.S.
\end{enumerate}

\subsection{K-12 Programs Examples and Resources}
\subsubsection{For Teachers}\hfill\\
While Particle Physics is often considered too advanced to be taught at K-12 level, it is not when brought at the right level, in the right context~\cite{k12_CP2C5} and with the right tools. The QuarkNet Data Activities Portfolio (DAP) is a compendium of activities designed for teachers to use with their students. The activities range from skill-building (Level 0) to independent research (Level 4). The activities are tied to specific standards including Next Generation Science Standards (NGSS), AP Physics standards and IB standards. Also, the DAP allows teachers to search the activities on the topics they cover such as Electricity \& Magnetism, Uncertainty, Conservation Law, etc. Teachers can find activities related to different experiments such as Cosmic Rays, LHC, and neutrinos. The activities have been vetted by teachers at QuarkNet Center meetings and by teachers and students in the classroom. Contact: Deborah Roudebush droudebush@cox.net
\\
\textbf{PhET Interactive Simulations:}
Founded in 2002 by Nobel Laureate Carl Wieman at the University of Colorado Boulder, PhET creates free interactive math and science simulations. The PhET project currently supports 21 simulations for quantum phenomena.  However, only two of these run on HTML5. Funds for translating the Java simulations into HTML5 simulations as well as developing new simulations is an opportunity to encourage teachers to include 21st Century ideas into the high school curriculum.
\\
\textbf{Cosmic Ray Studies:}
For over two decades QuarkNet has provided resources for faculty who wish to use cosmic ray studies in their outreach efforts. There are several levels of sophistication ranging from simple experiments performed virtually on existing data using complete analysis tools available to everyone, to operation of a high school detector within QuarkNet, to participation in large collaborative experimental efforts. 
\\
\textbf{Virtual Experiments:}
High schools and labs have collected 500k data files studying cosmic ray muons using QuarkNet scintillation detectors. That data and analysis tools are available to all users through an e-Lab. A document describing how to perform a set of simple experiments virtually (flux, lifetime, and muon speed) on a curated set of files is available at https://quarknet.org/content/resources-cosmic-ray-analyses-online.
\\
\textbf{Collaborative Projects:}
Groups of participating schools associated with multiple QuarkNet centers have organized larger research efforts that require collaboration. These cosmic ray measurements have led to student and teacher presentations at conferences and publications. Examples of these efforts include:
\begin{enumerate}

\item Cosmic ray muon rate during a total solar eclipse in 2017. Over 35 detectors collected data across the U.S. and improved the upper limit of muons from the sun detected at the earth’s surface by a factor greater than ten.

\item MUSE—cosmic ray rates at the NUMI underground neutrino beamline at Fermilab—measured the cosmic ray muon rate as a function of depth and of distance from the access shaft. The background of muons from neutrino interactions had to be subtracted.

\item Muon shadow cast by the moon—an ongoing effort to locate the muon shadow in the low energies at the earth’s surface.

\item A measurement of the muon g-2 using cosmic rays is under development.

 Contact Mark Adams adams@uic.edu
 
 \end{enumerate}
\textbf{“Cosmic Watch”:}
MIT developed a small, inexpensive cosmic ray scintillation detector that students can build. It is based on SiPM photon detection, read out by an Arduino; can operate in pairs to perform a variety of counting experiments. http://www.cosmicwatch.lns.mit.edu/ Contact Spencer Axani saxani@mit.edu.
\subsubsection{For Teachers and their Students}\hfill\\
\textbf{International Masterclasses:(IMC)} -
https://physicsmasterclasses.org - Physicists invite high school teachers to bring their students to universities or labs to be “particle physicists for a day.” Students learn about the standard model, problems in particle physics, and specific experiments from which they analyze authentic data. Students combine and share results. Teachers can act as auxiliaries to the masterclass mentors and tutors and, over time, become experts on the data. IMC happen in and around March each year, worldwide, with final discussions of results in video conferences hosted by CERN, Fermilab, KEK, GSI, or TRIUMF. IMC are organized by IPPOG, http://ippog.org/. IMC can be held online but in-person is best. Contact: Ken Cecire kenneth.w.cecire.1@nd.edu

Examples of Masterclasses: 
\begin{itemize}
\item ATLAS, CMS, ALICE, LHCb (LHC)
\item MINERvA Neutrino  Experiment (Fermilab)
\item Belle II (KEK)
\item Particle Therapy (GSI)
\item Darkside (INFN)
\end{itemize}

\subsubsection{For Students}\hfill\\
\textbf{Classroom Visits and Advanced Placement:}
At the K-12 level, role models~\cite{k12_CP2A1,k12_CP2A2,k12_CP2A3} take on more dynamics to demonstrate what is possible for the future. Students, especially young females, need to see and talk to scientists who look like them. Education literature confirms students need to hear how to get from their present grade level into a physics career. Advanced Placement programmes~\cite{k12_CP2C1} are a good way for students to find out what studies like physics are really about, which may be quite different from what they think based on the highschool curriculum. Across the country, particle physicists volunteer to speak to classrooms about what a physicist does and often work with a teacher to plan a physics activity around a curriculum topic. For example, Fermilab offers, Fermilab Science in Your Classroom,  https://ed.fnal.gov/home/virtualclassroomvisit.shtml Contact: Office of Education and Public Engagement Registrar, edreg@fnal.gov. 
\\
\textbf{Saturday Morning Physics:}
The Fermilab Saturday Morning Physics (SMP) program is a free-of-charge series of eleven lectures and tour visits given by Fermilab staff. SMP’s purpose is to further understanding and appreciation of modern physics among high school students. The lectures are two hours long, with a 10-minute break. Virtual tours and Q\&A sessions start at 11:00; with everything wrapping up around noon. SMP is held twice during each school year, once in fall and once in spring. Students must apply and be accepted to attend. It is helpful to have had a high school physics course and a couple of years of algebra. Fermilab encourages and welcomes parents and siblings to attend SMP on graduation day. There is a parallel Q\&A session for parents. Siblings may attend either the Q\&A or the final lecture (space permitting). High school teachers are welcome to attend, but only registered students may ask questions. SMP leaders are especially eager to hear, and to answer, teachers’ questions outside the lecture hall. Contact: Amanda Early aearly@fnal.gov
\\
\textbf{Space Explorers:}
A signature program of the Kavli Institute for Cosmological Physics, the Space Explorers Program offers inner-city youth from neighborhoods around the University of Chicago a multi-year science enrichment opportunity. The program provides over one-hundred contact hours each year including weekly laboratories taught on campus, three-day winter and week-long summer residential science institutes. This sustained engagement offers the Space Explorer participants the opportunity to become familiar with the university research community and the culture of science. It also helps to cultivate future teacher/scholars by offering a variety of valuable teaching, communication and team-working experiences to younger scientists. Participants are recruited by the Office of Special Programs, our partner with deep community roots, and are selected based on interest and commitment rather than grades or test scores. Each year between 20-30 Chicago Public middle and high school students participate, predominantly African-American with well over one half being female. On average, students stay in the program for three years, and the results are impressive. Graduates of Space Explorers matriculate in college as science majors at a rate that is 500\% better than is predicted by combined national and Chicago Public Schools statistics.  Over 50\% of Space Explorers graduates choose a STEM related field of study for their college degree.
\\
\textbf{Residential Science Institutes:}
The biannual residential institutes provide an immersive environment that encourages scientific curiosity and exploration for inner-city youth. A retreat-like setting offers ample time for in-depth studies (e.g., 5.5 hrs/day lab) and for valuable informal interactions with the scientists. During the first half of the institute, students explore thematically linked laboratories and investigations in small groups. During the latter part of the institute, each of three reporting groups explores a single laboratory in greater depth (eand prepares a presentation for parents and peers that culminate the institute. Contact: Brian Nord nord@fnal.gov

\subsection{Recommendations for K-12 Outreach Programs}
\begin{enumerate}
 \item HEP should form a K-12 particle physics outreach community with a “forum” for networking and sharing resources and open to physicists, engineers, technicians, computer scientists and secondary school educators. The community can discuss ideas and issues and work collaboratively to help develop understanding appropriate for secondary school about how scientists develop knowledge.
 \item Individuals can form local collaborative communities by Identifying a few teachers to hold an assessment session to determine the needs and interests of teachers. Then, the  physicists can find ways to meet needs such as helping teachers strengthen their scientific background and/or providing support to use instructional materials.
 \item K-12 programs should prioritize exposing all students to age appropriate STEM topics to create scientifically literate citizens and be mindful of bias in program designs that might filter student participation by race, gender, or socioeconomic background. 
 \item K-12 particle physics programs should engage with sister STEM fields with successful outreach programs to the same ages and participate in interdisciplinary discussions on best practices in science communication. 
 \item Funding agencies should provide funding for these outreach activities including compensation for those who organize and run them. 
 \item The community, universities, labs, and institutes should acknowledge and reward those who participate in this important work.
\end{enumerate}

\section{Opportunities in Undergraduate Education}

The Working Group discussed initiatives that support undergraduates as they begin their college years as a way to keep more students in physics. Also, members expressed concern that if particle physics does appear in the undergraduate curriculum, it is an upper level course. This may be one of the reasons that the rise in the number of physics BSc degrees lags behind the rise in STEM BSc degrees~\cite{k12_CP2C6}. We discussed incorporating particle physics ideas into the curriculum much earlier. Perhaps, a survey to QuarkNet mentors could measure the infusion of particle physics ideas throughout the undergraduate curriculum. Also, we noted a need for a project repository with recommended readings, successful programs, studies and reports.

\subsection{Program Examples}
\subsubsection{Math Tutorial for Freshmen}\hfill\\
Most undergraduate physics majors do not embark upon a research experience, if indeed at all, until they are juniors or seniors. By this time, many already must think about their path after college, whether it be in industry or academia. If the students have not had research experiences, particularly in physics areas such as particle physics or astrophysics, they would not be able to fully discern the best fit for them post graduation. On the other hand, it is crucial for students to begin to connect their physics understanding in the classroom with current research as early as possible. Therefore, one proposal that has been implemented in some universities is to introduce the fields of current physics research, including particle physics, to freshmen students via a project-based research experience~\cite{k12_CP2C4}. Not only will incoming physics majors be able to learn about careers and make a more informed decision about their future, but they can also start to apply what they learned in the classroom right away through tailored projects. Some skills that students would be exposed to include statistical reasoning, programming, and scientific writing and communications, all are useful skills that freshmen can easily learn in order to make meaningful contributions to research even before taking upper level physics coursework. 
\subsubsection{Freshman Projects}\hfill\\
Loyola University Chicago’s Department of Physics requires freshman physics majors to enroll in a one credit semester course called Freshman Projects. This class is a “project-based learning experience” that will introduce them to scientific research through designing a project with a Loyola physics faculty member. Students carry out an experimental and/or a theoretical analysis throughout the semester and present their findings in a presentation at the end of the term.  This course was created because, as described by the LUC Physics Department, “while undergraduate research has become an important part of many Physics departments, students frequently first become involved in research in their Junior or Senior year. By contrast, the Freshman Project models the research process by involving students during their first year in the program. This innovative project was started in the Spring Semester of 1996 by Dr. Asim Gangopadhyaya as a part of an introductory course for Physics majors, and since that time, the project has become its own 1-credit course that is required for all majors in our department.” 
This course has shown to give undergraduates experience in scientific writing and communication, broaden problem solving skills, and help to expose them to current topics in the field of physics. Also, many projects end up being incorporated into larger research work culminating in publications for some of the students. Contact Asim Gangopadhyaya agangop@luc.edu. 
\subsubsection{Physics for Future Presidents}\hfill\\
Physics for Future Presidents is a course, book, and publicly available lecture developed by Prof. Richard Muller at UC Berkeley under the premise that a conceptual understanding of modern physics is both essential and possible for tomorrow’s political leaders~\cite{k12_CP2C2}. The original course is offered as a lecture that fulfills the general education science requirement for non-science majors. This type of course is now offered by universities across the country as a freshman seminar style course to introduce non-science majors to the fundamentals of modern physics in the context of how they might interact with those topics in our modern society as voters or policy makers. This style of course often involves laboratory and writing components.  “A serious but accessible presentation of topics important for leaders to know—energy, global climate, high-tech devices, space travel, nuclear weapons, etc. Students possessing any level of scientific background are equipped with the concepts and analytical tools needed to make informed decisions and to argue their view persuasively” - http://bulletin.gwu.edu/courses/phys/
\subsubsection{Study Groups}\hfill\\
Many students at the undergraduate level would benefit from support groups and collaboration opportunities that would help them to matriculate effectively into their given physics program. Also, they would be a source of support throughout the student’s undergraduate journey. Not only would such groups be peer mentors who can give help academically, but they can also be havens of support and encouragement for underrepresented minority groups in STEM. Such a structure within a physics department could help to retain more physics majors, thus increasing the number that will successfully finish a Bachelor's degree. Combining this with reducing the class size of physics major courses for a more personal and intimate approach to learning from professors, this will untimely provide undergraduates with a strong overall support system that appreciates, accepts, and guides them to better success.
\subsubsection{Career Education}\hfill\\
The comprehensive report by the Joint Task Force on Undergraduate Physics Programs (J-TUPP), convened by the APS and the AAPT, is targeted toward helping physicists prepare undergraduates for a diverse set of careers. In particular, chapters five and seven offer suggestions to improve recruitment and retention of undergraduates into particle physics, including research experiences, career-focused workshops, and curriculum reform of courses in the physics major. While focused on undergraduate institutions, the J-TUPP recommendations parallel several aspects of QuarkNet’s work with high school teachers and students. A collaboration with undergraduate program leaders for training undergraduate educators in the use of inquiry-based instruction infused with particle physics could generate novel support for the undergraduate-to-researcher path.
The Physics Teacher Education Coalition (PhysTEC) is a program designed to address the national shortage of qualified high school physics teachers. Increasing the number of high school students with a solid preparation directly impacts the ability of undergraduate institutions to recruit and retain students. Particle physicists can begin participating in PhysTEC efforts by simply sharing its materials on teaching as a career option with their local undergraduate society.

\subsection{Recommendations for Undergraduate Programs}
\begin{enumerate}
\item Funding and recognition for individuals who support and develop a culture of inclusivity and equity in the department, including, but not limited to attending conferences on education and diversity, publication costs for education and DEI papers, mentoring and event planning 
\item In admission and orientation to physics programs it is important not to place a cutoff by which students who have not had calculus before college have no tenable pathway as a physics major. AIP statistics on math education in high school demonstrate a clear racial and socioeconommic disparity between students who have had calculus before graduating and those who have not, indicating that this level of achievement at this stage is not predictive of future achievement or mathematical prowess. The number of students able to access a physics degree could be tripled by the removal of this one requirement, allowing students to begin calculus as freshmen. 
\item Offer engaging general education physics courses, prioritizing topics of strategic interest to non-science students as scientifically literate citizens. While a physics student may need to master kinematics first, someone who is not going to pursue a physics degree does not. They need to form an understanding of modern physics that helps them navigate media and politics, and an appreciation for how scientists know what they claim to know. This can be accomplished with a modern physics first approach. 
\item Prioritize teaching methods that have been verified as effective in physics education, especially discovery learning, project-based coursework, and active learning techniques. Funding for classroom materials and professional development should be made available for this purpose. 
\item Where possible, ask students to start research projects early. Do not overemphasize publishability in undergraduate projects, repeating something that has been done, but is very instructive or foundational should be acceptable for the process of learning what research is and how to do it. 
\item Encourage students, especially minoritized and intersectionality minoritized students to have multiple mentors at the same time. Studies have shown this increases persistence in female graduate students and the result is likely more broadly applicable. 
\item Perhaps the forum described by Working Group 3 is a place for the repository. 
 \end{enumerate}
 
\section{Educational Opportunities for All}
Particle physics education efforts should strive for inclusivity, to foment interest in students across the social spectrum and provide them with the background, tools and opportunities to flourish. No individual or societal group has an a priori proxy for potential in academic achievement, research ability or professional accomplishments. 

It is therefore important to view outreach to K-12 and undergraduate education through the broad lens of equity and inclusivity. The HEP community should be mindful of potential instructor bias and self-selection effects that could reduce the impact of outreach opportunities on the eventual diversity within particle physics as a profession. 

Opportunities for undergraduate students to engage with physics and particle physics should similarly be designed to support and foster a culture of diversity, equity, and inclusion with the understanding that students will come into these programs with biases about themselves and their peers~\cite{k12_CP2A7}. The effectiveness of role models in fighting the undesired biases should be stressed, not only for teachers~\cite{k12_CP2A1,k12_CP2A2,k12_CP2A3}, but also in the curriculum \cite{k12_CP2A4}. It is also important to paint a good picture of career perspectives related to cultural and social settings \cite{k12_CP2A5,k12_CP2A6,k12_CP2A8,k12_CP2A9,k12_CP2A10}.

Recruiting students broadly from diverse backgrounds should be paired with equal attention to ensuring that the programs they are recruited into are also conducive to their success. Significant background material is available from a number of sources to help guide these  important programmatic efforts for the field, such as by STEP-UP a national community of physics teachers, researchers, and professional societies that developed and implement high school physics lessons to empower teachers, create cultural change, and inspire young women to pursue physics in college~\cite{k12_CP2A11}, and a report that provides actionable recommendations for community wide efforts to address the long-term systemic issues within the physics and astronomy communities that contribute to the underrepresentation of African Americans in those fields by the AIP~\cite{k12_CP2A12}.

Ongoing, evolving and new efforts to ascertain inclusivity in education at all levels need unwavering support, in particular also by the particle physics community.

\section{Conclusions}

This paper provides an overview of several existing programs at K-12 and undergraduate levels aimed to foster and develop increased student interest early on in the fundamental  sciences and  particle  physics, while simultaneously strengthening equity, diversity and inclusively in student and teacher participation. Guided by such efforts, recommendations are provided to inform and further the building of strong interconnections between the research community and K-12 and undergraduate level teachers and students, regarded as important for meaningful systemic educational change.

\section*{References}

\typeout{}
\bibliography{iopart-num}


\end{document}